\documentclass[a4paper,10pt,twoside]{cpc-hepnp}

\usepackage{multicol}
\usepackage{graphicx}
\usepackage{booktabs}
\usepackage{amssymb,bm,mathrsfs,bbm,amscd}
\usepackage[tbtags]{amsmath}
\usepackage{lastpage}

\begin{document}
%\begin{CJK*}{GBK}{song}

\fancyhead[c]{\small submitted to ¡¯Chinese Physics C¡¯}
 \fancyfoot[C]{\small 010201-\thepage}
\title{The Side-Effects of the Space Charge Field Introduced by Hollow Electron Beam in the Electron Cooler of CSRm\thanks  {Supported by National Natural Science
Foundation of China (11375245) }}

\author{%
      TANG Mei-Tang£¨ÌÀ÷Ìã©$^{1,2;1)}$\email{tangmt@impcas.ac.cn}%
\quad YANG Xiao-Dong£¨ÑîÏþ¶«£©$^{1}$
\quad MAO Li-Jun£¨Ã°Á¢¾ü£©$^{1}$
\quad LI  Jie£¨Àî½Ü£©$^{1}$
\quad MA Xiao-Ming£¨ÂíÏþÃ÷£©$^{1}$
\quad YAN Tai-Lai£¨êÌÌ«À´£©$^{1}$\\
\quad ZHENG Wen-Heng£¨Ö£Îĺࣩ$^{1,2}$
\quad ZHAO He£¨ÕԺأ©$^{1,2}$
\quad WU Bo£¨ÎⲨ£©$^{1,2}$
\quad WANG Geng£¨Íõ¹¢£©$^{1,2}$
\quad RUAN Shuang£¨Èîˬ£©$^{1,2}$
\quad SHA Xiao-Ping£¨É³Ð¡Æ½£©$^{1}$
}
\maketitle

\address{%
$^1$ Institute of Modern Physics, Chinese Academy of Sciences, Lanzhou 730000, People¡¯s Republic of China\\
$^2$ University of Chinese Academy of Sciences, Beijing 100049, People¡¯s Republic of China\\
}

\begin{abstract}
Electron cooler is used to improve the quality of the beam in synchrotron, however it also introduces nonlinear electromagnetic field, which cause tuneshift, tunespread and may drive resonances leading to beam loss. In this paper the tuneshift and the tunespread caused by nonlinear electromagnetic field of the hollow electron beam was investigated, and the resonance driving terms of the nonlinear electromagnetic field was analysed. The differences were presented comparing with the solid electron beam. The calculations were performed for $^{238}U^{32+}$ ions of energy 1.272MeV stored in CSRm, using the parameters given in table1. The conclusion is that in this situation nonlinear field caused by the hollow electron beam do not lead to serious resonances.
\end{abstract}



\footnotetext[0]{\hspace*{-3mm}\raisebox{0.3ex}{$\scriptstyle\copyright$}2013
Chinese Physical Society and the Institute of High Energy Physics
of the Chinese Academy of Sciences and the Institute
of Modern Physics of the Chinese Academy of Sciences and IOP Publishing Ltd}%

\begin{multicols}{2}

\section{Introduction}

Two electron cooler devices with the most important characteristic that the distribution of the electron beam is adjustable have been installed at the HIRFL-CSR. Generally, the hollow electron beam is used to cool the ion beam stored in CSR, so analyzing the side-effects of the hollow electron beam to ion beam is necessary for HIRFL-CSR to improve its beam quality. Because the space charge field of the electron beam has larger effects on lower-energy ions, the $^{238}U^{32+}$ ions of energy 1.272MeV/u are chosen as a typical example in the calculation. The paraments used in the calculation are summarised in table1.
\begin{center}
\tabcaption{ \label{tab2}  Parameters used in the calculations.}
\footnotesize
\begin{tabular*}{80mm}{@{\extracolsep{\fill}}c|c}
\toprule Particle & $^{238}U^{32+}$ 1.272MeV \\
\hline
Currents of hollow  & 0.077A\\
electron beam   &   \\
\hline
Currents of solid  & 0.077 A \\
electron beam &   \\
\hline
Parameters for the electron  & $a1=2.86*10^{-4},b1=14.3*10^{-2 }$\\
 beam distribution &   $a2=2.86*10^{-4},b2=28.6*10^{-2}$\\
\hline
Cooling length  Lcool & 4m\\
\hline
Beta function in the cooler &$\beta_x=10m,\beta_y=17m$\\
\hline
Tune of CSRm & $Q_x=3.63, Q_y=2.61m$\\
\bottomrule
\end{tabular*}%
\end{center}

The radial distribution of the electron beam in the cooler can be parameterized by follow equations[1], the equation (1) is for hollow electron beam and the equation (2) is for solid electron beam.
\begin{equation}\label{1}
  \rho \left( {\rm{r}} \right) = \frac{{{\rho _h}}}{2}\left( {{\rm{erfc}}\left( {{\rm{(r - }}{{\rm{b}}_{\rm{2}}}{\rm{)/}}{{\rm{a}}_{\rm{2}}}} \right) - {\rm{erfc}}\left( {{\rm{(r - }}{{\rm{b}}_{\rm{1}}}{\rm{)/}}{{\rm{a}}_{\rm{1}}}} \right)} \right).\
\end{equation}
\begin{equation}\label{2}
\rho \left( {\rm{r}} \right) = \frac{{{\rho _s}}}{2}{\rm{erfc}}\left[ {{\rm{(r - }}{{\rm{b}}_2}{\rm{)/}}{{\rm{a}}_2}} \right].\
\end{equation}

Where a1, a2, b1, b2 are constants determine the shape of the distribution(see table1), $\rho _h ,\rho _s$ are normalization coefficient. Fig.1 shows the radial distribution for solid electron beam and hollow electron beam.

\begin{center}
\includegraphics[width=7cm]{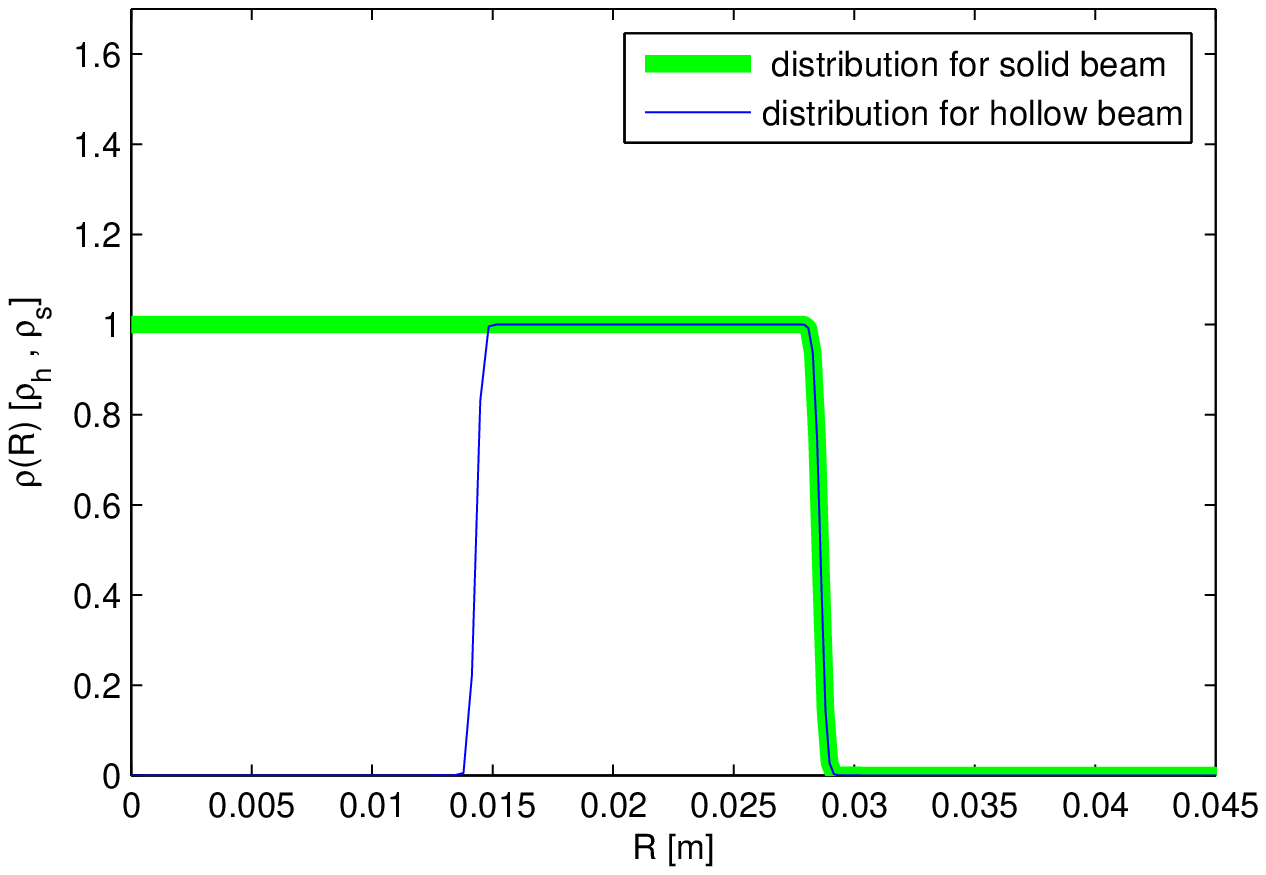}
\figcaption{\label{fig1}   The radial distribution of the electron beam.}
\end{center}

Because the longitudinal length of the electron beam is far larger than its transverse width, and the distribution of electron beam is axisymmetric. Using the Ampere's law and Gauss's law, the magnetic field and the electric field caused by the electron beam can be obtained:
\begin{equation}\label{3}
\vec E\left( R \right) = \frac{{\int\limits_0^R {\rho \left( r \right)rdr} }}{{R{\varepsilon _0}}}\hat r.\
\end{equation}
\begin{equation}\label{4}
\vec B\left( R \right) = \frac{{{\mu _0}\beta c\int\limits_0^R {\rho \left( r \right)rdr} }}{R}\hat \phi .\
\end{equation}
Where $\epsilon_0$ is the permittivity, $\mu_0$ is the permeability, $\beta$ is the relativistic factor, R is the transverse location, $\rho(r)$ is the radial distribution of the electron beam in equation (1), (2).
The particle has charge q at the location R will encounter the force:
\begin{equation}\label{5}
  F(R) = qE(R) + qv \times B(R).\
\end{equation}
Using the equation (2), (3)and (4), then the force become:
\begin{equation}\label{6}
 F(R) = \frac{q}{{R{\varepsilon _0}{\gamma ^2}}}\int\limits_0^R {\rho \left( r \right)} rdr.\
\end{equation}
Where q is the charge of the particle, $\gamma$ is the Lorentz factor. As fig.2 and equation (6) show the force that the particle encounter in the electron beam is nonlinear. So the tuneshift, tunespread will be caused and resonances may will be driven.
\begin{center}
\includegraphics[width=7cm]{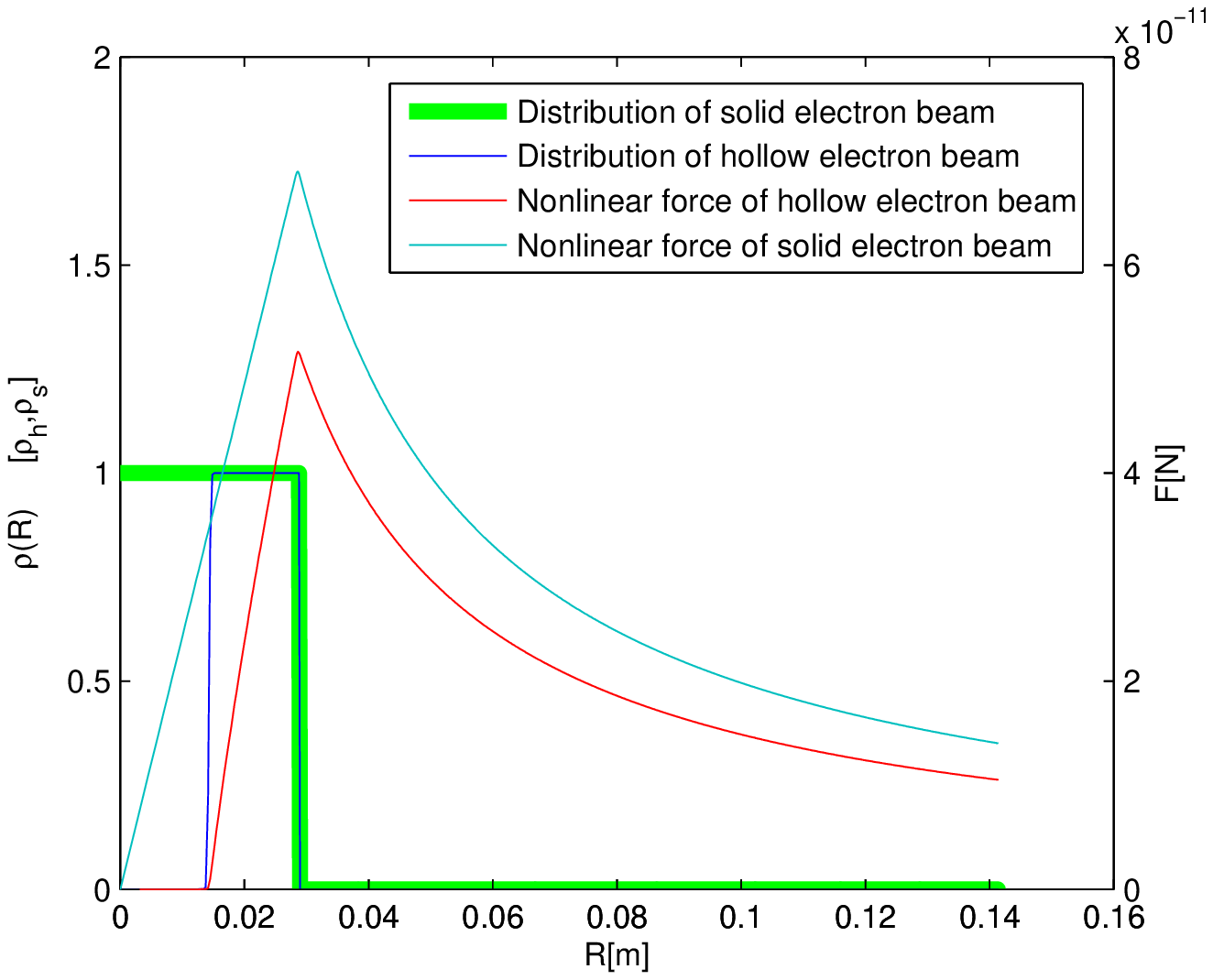}
\figcaption{\label{fig2} The nonlinear force exerts on the particles at different radius.}
\end{center}

In section 2 and 3 the calculations of tuneshift and tunespread for hollow electron beam and solid electron beam are presented separately, in section 4 the resonance driving terms are analyzed, and in section 5 the conclusions are summarised.
\section{Tuneshift}

With thin lens approximation, the nonlinear transverse kick can be obtained[2,3]:
\begin{equation}\label{7}
  \Delta r' = \frac{{qq'N'}}{{2\pi {\varepsilon _0}{\beta ^2}{m_0}{c^2}{\gamma ^3}R}}\frac{{\int\limits_0^R {r\rho \left( r \right)} dr}}{{\int\limits_0^\infty  {r\rho \left( r \right)} dr}}.\
\end{equation}
\begin{equation}\label{8}
N' = \left| {\frac{{{L_{cool}}{I_e}}}{{q'{\beta}c}}} \right|.\
\end{equation}
Where q' is the charge of electron, Lcool is the lengths of cooler section, $m_0$ is the rest mass of the particle, N' is the number of electrons in the electron cooler[4].
Imitating the calculating method in ref.2. The differential equation for the transverse motion with the kicker of the electron beam can be written as follow[2]:
\begin{equation}\label{9}
  \frac{{{d^2}z}}{{d{s^2}}} + K(s)z = \Delta z'\delta (s).\
\end{equation}
\begin{equation}\label{10}
  \Delta z' = \frac{{qq'N'}}{{2\pi {\varepsilon _0}{\beta ^2}{m_0}{c^2}{\gamma ^3}z}}\frac{{\int\limits_0^R {r\rho \left( r \right)} dr}}{{\int\limits_0^\infty  {r\rho \left( r \right)} dr}}.\
\end{equation}
Using the Courant-Snyder transformation
\begin{equation}\label{11}
  \eta  = z/\sqrt {{\beta _z}} ,\theta  = \int {ds/(Q{\beta _z})} .\
\end{equation}
the equation (9) will become:
\begin{equation}\label{12}
  \frac{{d{\eta ^2}}}{{d{\theta ^2}}} + {Q^2}\eta  = Q\sqrt {{\beta _z}} \Delta z'(\eta ,\theta )\delta (\theta ).\
\end{equation}
Where Q is the tune. Transform the equation to  $\varepsilon$ , $\phi$.
\begin{equation}\label{13}
  \eta  = \sqrt \varepsilon  \cos (\phi ),\eta ' = \sqrt \varepsilon  \sin (\phi ).\
\end{equation}
Replacing the period Dirac function in equal.(12) by its Fourier expansion.
\begin{equation}\label{14}
  \delta (\theta ) = \frac{1}{{2\pi }}\sum\nolimits_{ - \infty }^{ + \infty } {\exp ( - imp)}.\
\end{equation}
Then the equation of the transverse motion can be Transform to two first-odder differential equations:
\begin{equation}\label{15}
  \frac{{d\varepsilon }}{{d\theta }} = \frac{{\sqrt \varepsilon  \sin (\phi )\sqrt \beta  \Delta r'}}{\pi }\sum\limits_{m =  - \infty }^\infty  {\exp ( - im\theta )}.\
\end{equation}
\begin{equation}\label{16}
  \frac{{d\phi }}{{d\theta }} = Q + (\frac{{\sqrt \beta  \cos (\phi )\Delta r'}}{{2\pi \sqrt \varepsilon  }})\sum\limits_{m =  - \infty }^\infty  {\exp ( - im\theta )}. \
\end{equation}
The nonlinear tuneshift can now be calculated by averaging the right hand side of equation (16)
over $\theta$ and $\phi$:
\begin{equation}\label{17}
  \Delta \nu  = \frac{{qq'N}}{{8{\pi ^3}\varepsilon {\beta ^2}{c^2}{m_0}{\varepsilon _0}{\gamma ^3}}}\int\limits_0^{2\pi } {\frac{{\int\limits_0^{\sqrt {{\beta _z}\varepsilon } \cos (\phi )} {r\rho (r)dr} }}{{\int\limits_0^\infty  {r\rho (r)dr} }}d\phi }.\
\end{equation}
Where $\beta_z$ instead the $\beta_x$ and $\beta_y$. Calculating equation (17) with different $\varepsilon$, the variations of tuneshift with the transverse oscillating amplitude of the particle($R=sqrt(\beta_z\epsilon)$) will be obtained.

Using the paraments in table1, the tuneshift can be worked out. The maximum tuneshift for hollow electron beam is $\Delta\nu_x=0.015,\Delta\nu_y=0.018$ and for solid electron beam is $\Delta\nu_x=0.024,\Delta\nu_y=0.041$. The variation of the  tuneshift with transverse oscillating amplitude of the particle is illustrated in fig.3.

\begin{center}
\includegraphics[width=7cm]{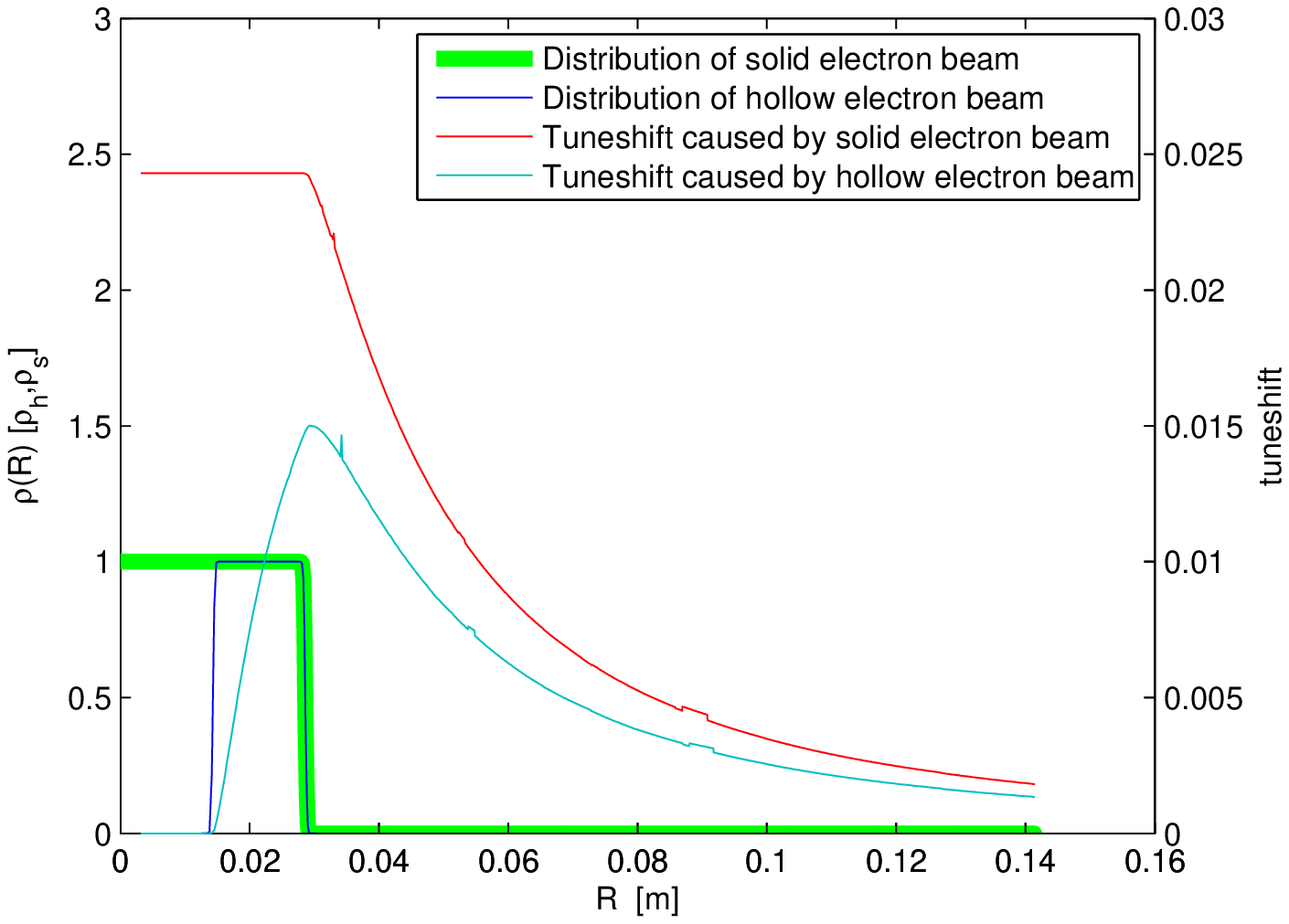}
\figcaption{\label{fig3}   The variation of tuneshift caused by electron beam (Just for the horizontal direction ) }
\end{center}

From fig.3 the follow phenomena can be observed:

1)The tuneshift caused by the hollow electron beam is smaller than the tuneshift caused by the solid electron beam.

2)For hollow electron beam the tuneshift is zero when the transverse oscillating amplitude of the particle is smaller than the inner radius of the hollow electron beam, while for the solid electron beam the tuneshift is a constant when the transverse oscillating amplitude of the particle is smaller than the range of the electron beam.

3)Once the oscillating amplitude of the particle is larger than the range of electron beam, the both tuneshifts are decrease with the increase of the transverse oscillating amplitude of the particle.

%The explanations for the above phenomena: 1)The nonlinear force caused by hollow electron beam is smaller than which caused by solid electron beam. 2)For the hollow electron beam the force is equal to 0 when the transverse oscillating amplitude of the particle is smaller than the inner radius of the hollow electron beam, but for the solid electron beam the force is linear when transverse oscillating amplitude of the particle is small enough[1]. 3)The magnitude of the nonlinear force is decrease with the increase of the transverse displacement of the particle when the particle is outside the electron beam.

\section{Tunespread}
From the equation (5) the two-dimensional equations of transverse motion can be gotten:
\begin{equation}\label{18}
  \frac{{d{\eta _x}^2}}{{d{\theta ^2}}} + {Q_x}^2{\eta _x} = {Q_x}\sqrt {{\beta _x}} \Delta x'\delta (\theta ).\
\end{equation}
\begin{equation}\label{19}
  \frac{{d{\eta _y}^2}}{{d{\theta ^2}}} + {Q_y}^2{\eta _y} = {Q_y}\sqrt {{\beta _y}} \Delta y'\delta (\theta ).\
\end{equation}

\begin{equation}\label{20}
  \Delta x' = \frac{{qq'N'x}}{{2\pi {\varepsilon _0}{\beta ^2}{m_0}{c^2}{\gamma ^3}{R^2}}}\frac{{\int\limits_0^R {r\rho \left( r \right)} dr}}{{\int\limits_0^\infty  {r\rho \left( r \right)} dr}}.\
\end{equation}

\begin{equation}\label{21}
  \Delta y' = \frac{{qq'N'y}}{{2\pi {\varepsilon _0}{\beta ^2}{m_0}{c^2}{\gamma ^3}{R^2}}}\frac{{\int\limits_0^R {r\rho \left( r \right)} dr}}{{\int\limits_0^\infty  {r\rho \left( r \right)} dr}}.\
\end{equation}

\begin{equation}\label{22}
  x = \sqrt {{\beta _x}{\varepsilon _x}} \cos ({\phi _x}).\
\end{equation}
\begin{equation}\label{23}
R = \sqrt {{\beta _x}{\varepsilon _x}{{\cos }^2}({\phi _x}) + {\beta _y}{\varepsilon _y}{{\cos }^2}({\phi _y})}. \
\end{equation}
Using the method in ref.4, replacing the period Dirac function in equation (18), (19) by its Fourier expansion:
\begin{equation}\label{24}
  \delta (\theta ) = \frac{1}{{2\pi }}\sum\nolimits_{ - \infty }^{ + \infty } {\exp ( - imp)}. \
\end{equation}
Only keeping the m=0 term of the Dirac function and averaging over phases $\phi_x $ and $\phi_y$, following equations will be obtained:
\begin{multline}\label{25}
  \Delta {\nu _x}({\varepsilon _x},{\varepsilon _y}) = \frac{1}{{4{\pi ^2}}} \\
  \int\limits_0^{2\pi } {d{\phi _y}\int\limits_0^{2\pi } {\frac{{{\beta _x}{{\cos }^2}({\phi _x})}}{{2\pi \sqrt {{\varepsilon _x}{\beta _x}{{\cos }^2}({\phi _x}) + {\varepsilon _y}{\beta _y}{{\cos }^2}({\phi _y})} }}} } d{\phi _x}.\
\end{multline}
\begin{multline}\label{26}
  \Delta {\nu _y}\left( {{\varepsilon _x},{\varepsilon _y}} \right) = \frac{1}{{4{\pi ^2}}} \\
  \int\limits_0^{2\pi } {d{\phi _y}\int\limits_0^{2\pi } {\frac{{{\beta _y}{{\cos }^2}({\phi _y})}}{{2\pi \sqrt {{\varepsilon _x}{\beta _x}{{\cos }^2}({\phi _x}) + {\varepsilon _y}{\beta _y}{{\cos }^2}({\phi _y})} }}} } d{\phi _x}.\
\end{multline}

Integrating the above integral with different $\varepsilon _x$ , $\varepsilon _y$, the $\Delta{\nu_x}$, $\Delta{\nu_y}$ can be worked out.

Using the above method and with paraments listed in table1, the tunespread can be obtained, as fig.4, fig.5 show.
\begin{center}
\includegraphics[width=7cm]{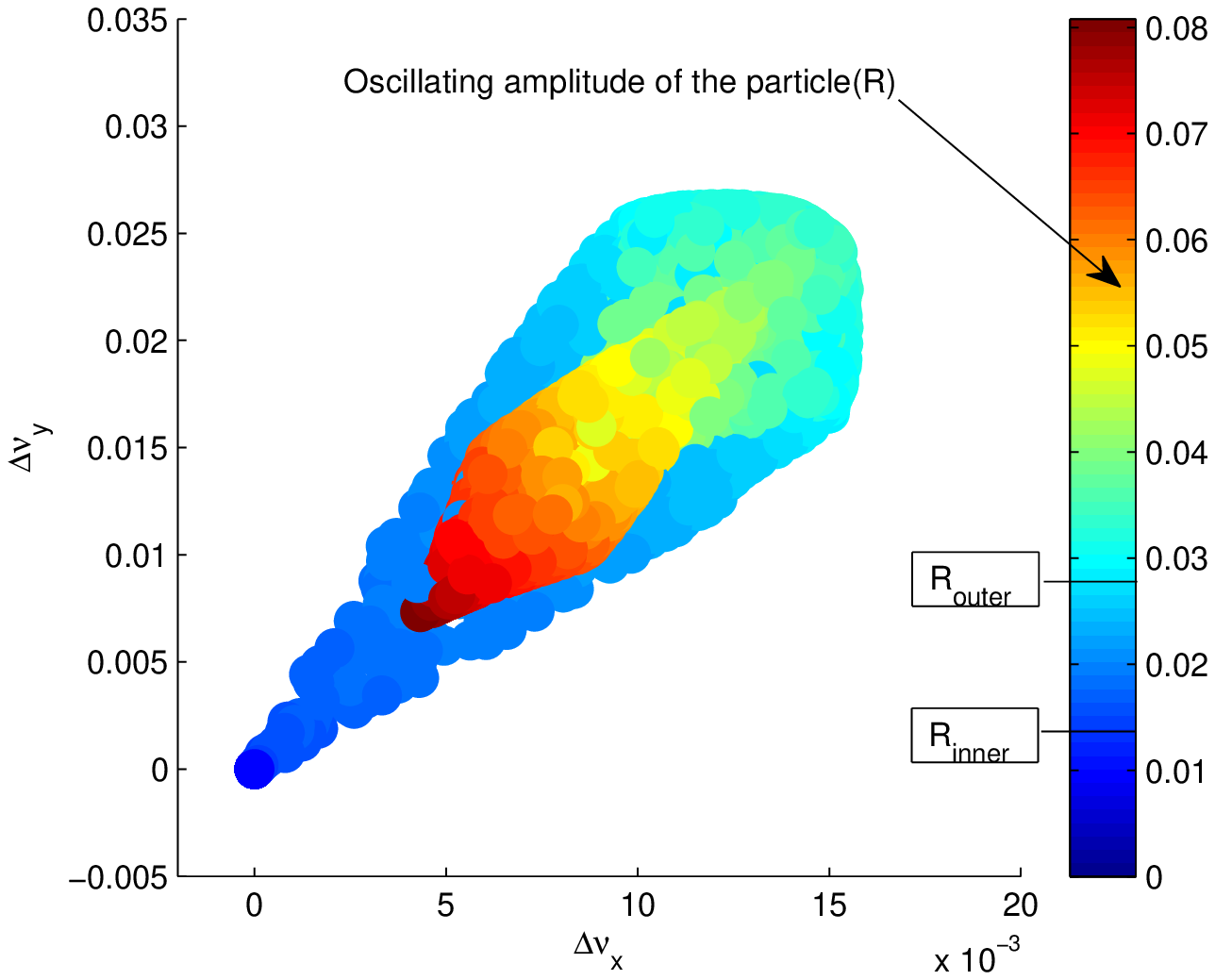}
\figcaption{\label{fig4} Tunespread caused by the hollow electron beam. $\varepsilon _z$ separately take the random number from 0 to ${(2*b2)^2}/{\beta _z}$ in the calculation. $R_{inner}, R_{outer}$ represent the ranges of the hollow electron beam.  }
\end{center}

Fig.4. shows that $\Delta{\nu_x}$ , $\Delta{\nu_y}$ stay close to 0, when the transverse oscillating amplitude of the particle is smaller than the inner radius of the hollow electron beam ($R_{particle} < R_{inner-radius}$), the bluest point at the lower left corner shows. When the transverse oscillating amplitude of the particle is not larger than the range of the electron beam($R_{inner-radius} <R_{particle} < R_{outer-radius}$), the $\Delta{\nu_x}$ , $\Delta{\nu_y}$  increasing along with the increasing of transverse oscillating amplitude of the particle. When the transverse oscillating amplitude of the particle is larger than the range of the electron beam($R_{particle} > R_{outer-radius}$), with the increasing of the transverse oscillating amplitude of the particle, the $\Delta{\nu_x}$ , $\Delta{\nu_y}$  decreasing.
\begin{center}
\includegraphics[width=7cm]{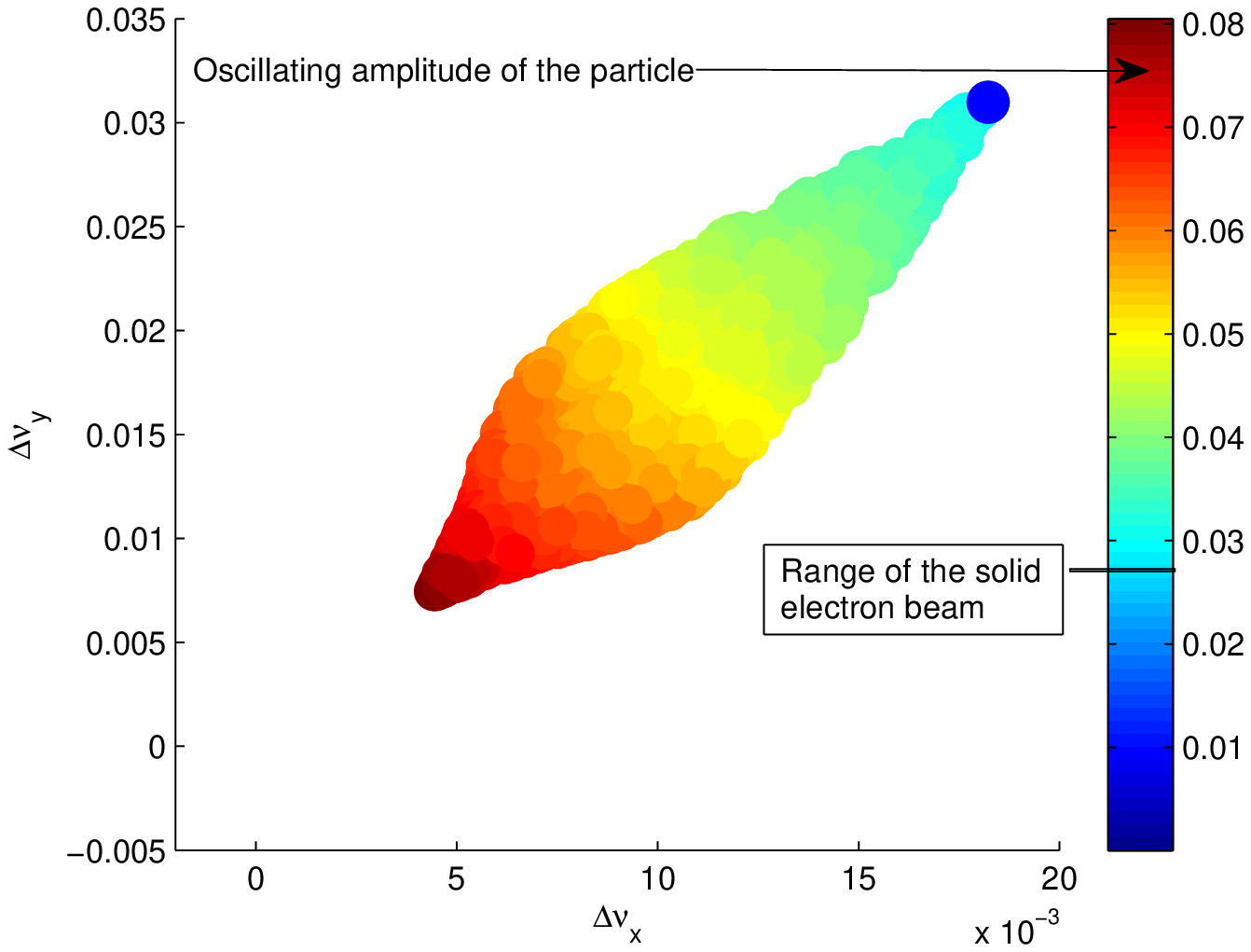}
\figcaption{\label{fig5} Tunespread caused by the solid electron beam. $\varepsilon _z$ separately take the random number from 0 to ${(2*b2)^2}/{\beta _z}$ in the calculation.}
\end{center}

Fig.5 shows that the particle has a largest tune shift, when the transverse oscillating amplitude of the particle is smaller than the range of the electron beam, as the bluest point at the top right corner shows. When the transverse oscillating amplitude of the particle become larger, the tuneshift $\Delta{\nu_x}$ , $\Delta{\nu_y}$ become smaller.

Comparing the tunespread in fig.4 and fig.5 follow conclusions can be drawn: The area of tunespread caused by the hollow electron beam is larger than it caused by solid electron beam, while the maximum of the tuneshift caused by the hollow electron beam is smaller than which caused by solid electron beam.
\section{Resonances}
Beside causing the tuneshift and the tunespread, the nonlinear force of the electron beam also result in resonances. To calculate the resonance driving terms, following the calculation of ref.2 again[1]. Replacing the part in the bracket of the equation (16) by its Fourier transform:
\begin{equation}\label{27}
  \frac{{\sqrt \beta  \cos (\phi )\Delta r'}}{{2\pi \sqrt \varepsilon  }} = \sum\limits_0^\infty  {{A_n}\cos (n\phi )}. \
\end{equation}
After some algebra, the follow expression can be obtain:
\begin{equation}\label{28}
  \frac{{d\phi }}{{d\theta }} = Q + \sum\limits_{n = 0}^\infty  {{A_n}} \sum\limits_{m =  - \infty }^\infty  {\cos (n\phi  - m\theta )}. \
\end{equation}
$A_n$ is the Fourier coefficient
\begin{multline}\label{29}
  A_n = \frac{1}{\pi }\int_{ - \pi }^\pi  {\frac{{eqN'cos(n\phi )}}{{4{\pi ^2}\varepsilon_0 {\beta ^2}{c^2}{m_0}{\varepsilon _0}{\gamma ^3}}}}
  [\frac{{\int_0^{^{\sqrt {\beta \varepsilon_z } \cos (\phi )}} {r\rho(r)dr} }}{{\int_0^\infty  {r\rho(r)dr} }}]d\phi. \
\end{multline}

$A_n$ is the resonance width defined in ref.2 and ref.3, which is the function of $\varepsilon_z$ related to the transverse oscillating amplitude of the particle.

Integrating right side of the equation (29) numerically with paraments in table 1, and the $A_n$ can be worked out. Fig.7 shows the variations of $log_{10}(A_n)$ with the transverse oscillating amplitude of the particle.

\begin{center}
\includegraphics[width=7cm]{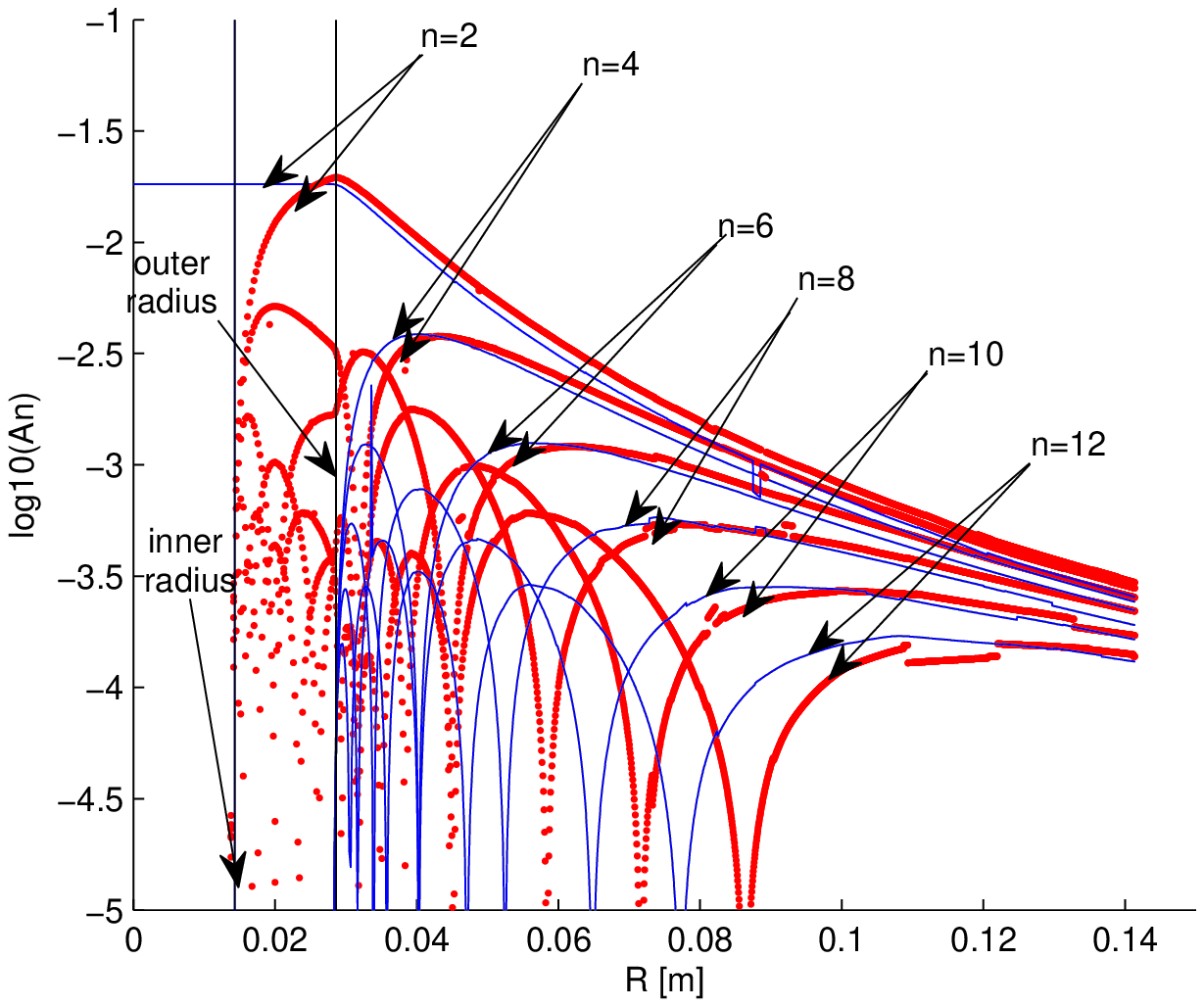}
\figcaption{\label{fig6} Blue lines show the resonance width for the solid electron beam. Red dotted lines show the resonance width for the hollow electron beam. R is the transverse oscillating amplitude of the particle. The black vertical lines are  the ranges of the electron beam.  }
\end{center}

The phenomena observed from the fig.6 are summarized as follow:

 1) Even order resonances have been driven by both the hollow beam and the solid beam.

 2) When the transverse oscillating amplitude of the particle is smaller than the inner radius of the hollow electron beam, the
    resonance width is zero. In other words, the resonance never occur when the particle is moving in the hollow part of the
    electron beam.

 3) For solid electron beam, when the transverse oscillating amplitude of the particle is smaller than the range of the electron beam, the resonance width for n=4,6,8,10,12 are equal to zero, and the resonance width for n=2 is a constant.

 4) With the increase of the resonance order, the resonance width for both electron beam are decreasing.

 5) When the transverse oscillating amplitude of the particle is large, the resonance width of the hollow electron beam and the
    solid electron beam are nearly same.

To know whether the tunespread caused by hollow electron beam calculated for $^{238}U^{32+}$ is large enough to cross some resonance-lines, plotting the resonance-lines on the tunespread figure (fig.7).
\begin{center}
\includegraphics[width=7cm]{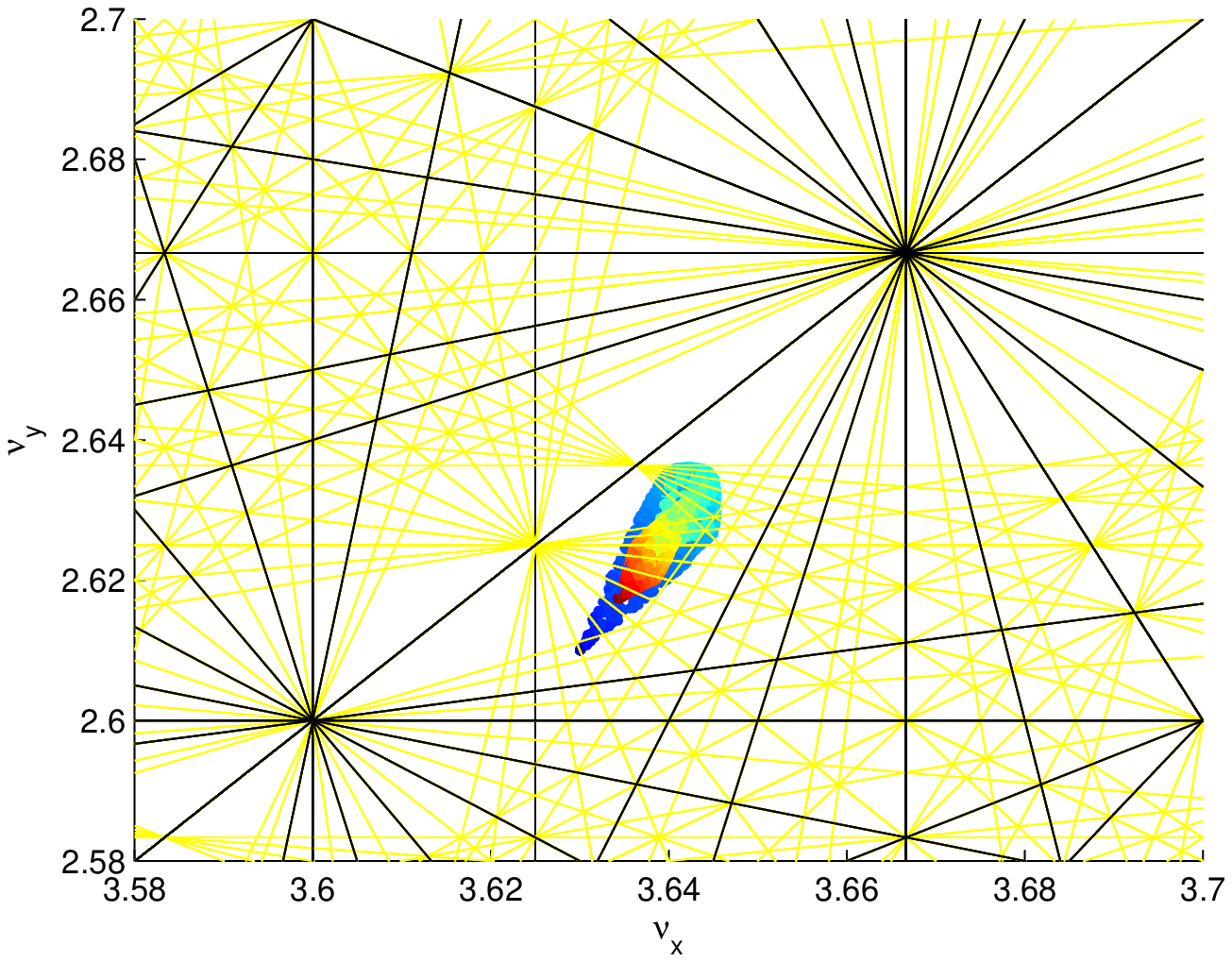}
\figcaption{\label{fig7} The yellow lines represent resonance-lines of order equal to 8,9,10,11,12; The black lines represent resonance-lines of order smaller than 8.}
\end{center}

As fig.7 shows, the tunespread just cross the resonance-lines of order larger than 7. In other words, under the conditions defined in table 1 nonlinear field caused by the hollow electron beam do not lead to resonances order smaller than 7.

\section{Conclusions}
 The tuneshift and tunespread caused by the hollow electron beam have been calculated, as a comparison the tuneshift and tunespread caused by the solid electron beam have also been calculated. The resonance driving terms for space charge field of the hollow electron beam and the solid electron beam have be obtained. As example, the calculation is performed for the beam $^{238}U^{32+}$ ions of energy 1.272MeV/u. The main conclusions are summarised: 1) The tuneshift caused by the solid electron beam is larger than which caused by hollow electron beam. 2) The area of the tunespread caused by hollow electron beam is larger than which caused by solid electron beam. 3) The even order resonances will be driven both by the hollow electron bean and the solid electron beam when the tune hits the resonance lines. 4)Under the conditions defined in table 1, the resonances that the order smaller than 7 do not happen for $^{238}U^{32+}$ ions of energy 1.272MeV/u in CSRm.

 Now the tuneshift, tunespread and the resonance driving terms caused by electron beam are clear. To further research the effects of the space charge field of the electron beam, the increase of the width and the emittance of ion beam due to the  resonances should be studied, and the tuneshift and tunespread caused by the space charge field of ion beam should also be analysed. It will be the work in the future.

\vspace{-1mm}
\centerline{\rule{30mm}{0.4pt}}

\end{multicols}

\clearpage

%\end{CJK*}
\end{document}